\documentclass[singlespacing]{bmcart}

\usepackage{graphicx}
\usepackage{bm}
\usepackage{enumitem}
\usepackage{url}

\usepackage{todonotes}

\usepackage{titletoc}
\usepackage{amsmath,amssymb}
\usepackage{hyperref}
\hypersetup{pdftitle={MAQRO - BPS 2023 Research Campaign Whitepaper}, breaklinks=true, colorlinks = true, urlcolor = blue, linkcolor = violet, frenchlinks = true, pdfborder = {0 0 0}}
\usepackage{soul}

\begin{document}

\begin{frontmatter}

\begin{fmbox}
\dochead{BPS2023 Research Campaign Whitepaper}


\title{Research Campaign: Macroscopic Quantum Resonators (MAQRO)}

\author[
	addressref={fmf,iqoqi},
	noteref={primary},
	email={rainer.kaltenbaek@fmf.uni-lj.si}
]{\inits{RK}\fnm{Rainer} \snm{Kaltenbaek}}
\author[
	addressref={univie},
]{\inits{MA}\fnm{Markus} \snm{Arndt}}
\author[
	addressref={iqoqi,univie},
]{\inits{MA}\fnm{Markus} \snm{Aspelmeyer}}
\author[
	addressref={ucl},
]{\inits{PFB}\fnm{Peter F.} \snm{Barker}}
\author[
	addressref={trieste,INFNtrieste},
]{\inits{AB}\fnm{Angelo} \snm{Bassi}}
\author[
	addressref={swansea},
]{\inits{JB}\fnm{James} \snm{Bateman}}
\author[
	addressref={tueb,QUB},
]{\inits{AB}\fnm{Alessio} \snm{Belenchia}}
\author[
	addressref={Onera},
]{\inits{JB}\fnm{Joel} \snm{Berg\'{e}}}
\author[
	addressref={MicroUlm,IQTulm},
]{\inits{CB}\fnm{Claus} \snm{Braxmaier}}
\author[
	addressref={ucl},
]{\inits{SB}\fnm{Sougato} \snm{Bose}}
\author[
	addressref={Onera},
]{\inits{BC}\fnm{Bruno} \snm{Christophe}}
\author[
	addressref={thorlabs,crystal},
]{\inits{GDC}\fnm{Garrett D.} \snm{Cole}}
\author[
	addressref={INFNfrascati},
]{\inits{CC}\fnm{Catalina} \snm{Curceanu}}
\author[
	addressref={warwick},
]{\inits{AD}\fnm{Animesh} \snm{Datta}}
\author[
	addressref={iqoqi},
]{\inits{MD}\fnm{Maxime} \snm{Debiossac}}
\author[
	addressref={univie},
]{\inits{UD}\fnm{Uro\v{s}} \snm{Deli\'{c}}}
\author[
	addressref={wigner,eotvos},
]{\inits{LD}\fnm{Lajos} \snm{Di\'{o}si}}
\author[
	addressref={NW},
]{\inits{AAG}\fnm{Andrew A.} \snm{Geraci}}
\author[
	addressref={univie},
]{\inits{SG}\fnm{Stefan} \snm{Gerlich}}
\author[
	addressref={brossel},
]{\inits{CG}\fnm{Christine} \snm{Guerlin}}
\author[
	addressref={adsF},
]{\inits{GH}\fnm{Gerald} \snm{Hechenblaikner}}
\author[
	addressref={brossel},
]{\inits{AH}\fnm{Antoine} \snm{Heidmann}}
\author[
	addressref={ZARM},
]{\inits{SH}\fnm{Sven} \snm{Herrmann}}
\author[
	addressref={duisburg},
]{\inits{KH}\fnm{Klaus} \snm{Hornberger}}
\author[
]{\inits{UJ}\fnm{Ulrich} \snm{Johann}}
\author[
	addressref={univie},
]{\inits{NK}\fnm{Nikolai} \snm{Kiesel}}
\author[
	addressref={ZARM},
]{\inits{CL}\fnm{Claus} \snm{L\"{a}mmerzahl}}
\author[
	addressref={nist},
]{\inits{TWL}\fnm{Thomas W.} \snm{LeBrun}}
\author[
	addressref={brisbane},
]{\inits{GJM}\fnm{Gerard J.} \snm{Milburn}}
\author[
	addressref={kcl},
]{\inits{JM}\fnm{James} \snm{Millen}}
\author[
	addressref={jpl},
]{\inits{MM}\fnm{Makan} \snm{Mohageg}}
\author[
	addressref={wright},
]{\inits{DCM}\fnm{David C.} \snm{Moore}}
\author[
	addressref={warwick},
]{\inits{GWM}\fnm{Gavin W.} \snm{Morley}}
\author[
	addressref={siegen},
]{\inits{SN}\fnm{Stefan} \snm{Nimmrichter}}
\author[
	addressref={ethPL,ethQC},
]{\inits{LN}\fnm{Lukas} \snm{Novotny}}
\author[
	addressref={SUPA},
]{\inits{DKLO}\fnm{Daniel K. L.} \snm{Oi}}
\author[
	addressref={QUB},
]{\inits{MP}\fnm{Mauro} \snm{Paternostro}}
\author[
	addressref={NTT},
]{\inits{CJR}\fnm{C. Jess} \snm{Riedel}}
\author[
	addressref={Onera},
]{\inits{MR}\fnm{Manuel} \snm{Rodrigues}}
\author[
	addressref={saclay},
]{\inits{LR}\fnm{Lo\"{i}c} \snm{Rondin}}
\author[
	addressref={IQTulm},
]{\inits{AR}\fnm{Albert} \snm{Roura}}
\author[
	addressref={IQTulm,IQSTulm,Hagler,AuMagri},
]{\inits{WPS}\fnm{Wolfgang P.} \snm{Schleich}}
\author[
	addressref={IQTulm},
]{\inits{TS}\fnm{Thilo} \snm{Schuldt}}
\author[
	addressref={duisburg},
]{\inits{BAS}\fnm{Benjamin A.} \snm{Stickler}}
\author[
	addressref={southampton},
]{\inits{HU}\fnm{Hendrik} \snm{Ulbricht}}
\author[
	addressref={ZARM},
]{\inits{CV}\fnm{Christian} \snm{Vogt}}
\author[
	addressref={IQTulm},
]{\inits{LW}\fnm{Lisa} \snm{W\"{o}rner}}

\address[id=fmf]{
  \orgname{University of Ljubljana, Faculty of Mathematics and Physics},
  \cny{Slovenia}
  }

\address[id=iqoqi]{
  \orgname{Institute for Quantum Optics and Quantum Information - Vienna},
  \cny{Austria}
  }
  
\address[id=univie]{
  \orgname{University of Vienna, Faculty of Physics, Vienna Center for Quantum Science and Technology},
  \cny{Austria}
  }
  
\address[id=ucl]{
  \orgname{Department of Physics and Astronomy, University College London},
  \cny{UK}
  }
  
\address[id=trieste]{
  \orgname{Department of Physics, University of Trieste},
  \cny{Italy}
  }

\address[id=INFNtrieste]{
  \orgname{INFN, Trieste Section},
  \cny{Italy}
  }
  
\address[id=swansea]{
  \orgname{Department of Physics, Faculty of Science and Engineering, Swansea University},
  \cny{UK}
  }
  
\address[id=tueb]{
  \orgname{Institut f\"{u}r Theoretische Physik, Eberhard-Karls-Universit\"{a}t T\"{u}bingen},
  \cny{Germany}
  }
  
\address[id=QUB]{
  \orgname{CTAMOP, School of Mathematics and Physics, Queen’s University Belfast},
  \cny{UK}
  }
  
\address[id=Onera]{
  \orgname{DPHY, ONERA, Universit\'{e} Paris Saclay},
  \cny{France}
  }
  
\address[id=MicroUlm]{
  \orgname{Institute of Microelectronics, Ulm University},
  \cny{Germany}
  }
  
\address[id=IQTulm]{
  \orgname{Institute of Quantum Technologies, German Aerospace Center (DLR)},
  \city{Ulm},
  \cny{Germany}
  }
  
\address[id=thorlabs]{
  \orgname{Thorlabs Crystalline Solutions},
  \cny{USA}
  }

\address[id=crystal]{
  \orgname{Crystalline Mirror Solutions, Santa Barbara, CA and Vienna},
  \cny{Austria}
  }
  
\address[id=INFNfrascati]{
  \orgname{INFN, Laboratori Nazionali di Frascati, Frascati (Roma)},
  \cny{Italy}
  }
  
\address[id=warwick]{
  \orgname{Department of Physics, University of Warwick},
  \cny{UK}
  }

\address[id=wigner]{
  \orgname{Wigner Research Center for Physics},
  \cny{Hungary}
  }
  
\address[id=eotvos]{
  \orgname{Department of Physics of Complex Systems, E\"{o}tv\"{o}s Lor\'{a}nd University},
  \cny{Hungary}
  }
  
\address[id=NW]{
  \orgname{Center for Fundamental Physics, Department of Physics and Astronomy,
Northwestern University},
  \cny{USA}
  }
  
\address[id=brossel]{
  \orgname{Laboratoire Kastler Brossel, Sorbonne Universit\'{e}, CNRS, ENS-Universit\'{e} PSL, Coll\`{e}ge de France},
  \cny{France}
  }
  
\address[id=adsF]{
  \orgname{Airbus Defence and Space GmbH, Friedrichshafen},
  \cny{Germany}
  }
  
\address[id=ZARM]{
  \orgname{ZARM, University of Bremen},
  \cny{Germany}
  }
  
\address[id=duisburg]{
  \orgname{Faculty of Physics, University of Duisburg-Essen},
  \cny{Germany}
  }
  
\address[id=nist]{
  \orgname{National Institute of Standards and Technology},
  \cny{USA}
  }
  
\address[id=brisbane]{
  \orgname{Centre for Engineered Quantum Systems, School of Mathematics and Physics,
The University of Queensland},
  \cny{Australia}
  }
  
\address[id=kcl]{
  \orgname{Department of Physics, King's College London},
  \cny{UK}
  }
  
\address[id=jpl]{
  \orgname{Jet Propulsion Laboratory, California Institute of Technology},
  \cny{USA}
  }
  
\address[id=wright]{
  \orgname{Wright Laboratory, Department of Physics, Yale University},
  \cny{USA}
  }
  
\address[id=siegen]{
  \orgname{Naturwissenschaftlich-Technische Fakult\"{a}t, Universit\"{a}t Siegen},
  \cny{Germany}
  }
  
\address[id=ethPL]{
  \orgname{Photonics Laboratory, ETH Z\"{u}rich},
  \cny{Switzerland}
  }
  
\address[id=ethQC]{
  \orgname{Quantum Center, ETH Z\"{u}rich},
  \cny{Switzerland}
  }
  
\address[id=SUPA]{
  \orgname{SUPA Department of Physics, University of Strathclyde},
  \cny{UK}
  }
  
\address[id=NTT]{
  \orgname{Physics \& Informatics Laboratories, NTT Research,\ Inc.},
  \cny{USA}
  }
  
\address[id=saclay]{
  \orgname{Universit\'{e} Paris-Saclay, CNRS, ENS Paris-Saclay, Centrale Sup\'{e}lec, LuMIn},
  \cny{France}
  }
  
\address[id=IQSTulm]{
  \orgname{IQST, University of Ulm},
  \cny{Germany}
  }

\address[id=Hagler]{
  \orgname{Hagler Institute for Advanced Study and Department of Physics and Astronomy, IQSE, Texas A\& M University},
  \cny{USA}
  }

\address[id=AuMagri]{
  \orgname{Texas A\& M AgriLife Research, Texas A\& M University},
  \cny{USA}
  }
  
\address[id=southampton]{
  \orgname{Department of Physics and Astronomy, University of Southampton},
  \cny{UK}
  }
  
\begin{artnotes}
\note[id=primary]{Primary White Paper Author. Names after the ﬁrst author sorted alphabetically} 
\end{artnotes}

\end{fmbox}


\begin{abstractbox}

\begin{abstract} 
The objective of the proposed MAQRO mission is to harness space for achieving long free-fall times, extreme vacuum, nano-gravity, and cryogenic temperatures to test the foundations of physics in macroscopic quantum experiments. This will result in the development of novel quantum sensors and a means to probe the foundations of quantum physics at the interface with gravity. Earlier studies showed that the proposal is feasible but that several critical challenges
remain, and key technologies need to be developed. These new technologies will open up the potential for achieving additional science objectives. The proposed research campaign aims to advance the state of the art and to perform the first macroscopic quantum experiments in space. Experiments on the ground, in micro-gravity, and in space will drive the proposed research campaign during the current decade to enable the implementation of MAQRO within the subsequent decade.
\parttitle{Primary area} The effects of the spaceflight environment physical systems and processes.
\end{abstract}

\begin{keyword}
 \kwd{quantum physics}
 \kwd{optomechanics}
 \kwd{matter waves}
 \kwd{optical trapping}
 \kwd{decoherence}
\end{keyword}


\end{abstractbox}

\smallskip

$^{\ast}$Primary White Paper Author. The other authors are sorted alphabetically.
$^{\dagger}$Correspondence: \href{mailto:rainer.kaltenbaek@fmf.uni-lj.si}{rainer.kaltenbaek@fmf.uni-lj.si}, phone: +43 (0) 664 1561372
%

\end{frontmatter}
\printaddresses%

\section{\label{sec::ScienceCase}Science Case and Motivation}
Are there fundamental limits to the size, complexity, or mass of quantum superpositions? Do we fully understand all fundamental sources of decoherence leading to the decay of macroscopic quantum states or will we see deviations due to yet unknown physics? Gravitational time dilation~\cite{Pikovski2015a,Zych2017a} or space-time fluctuations~\cite{Karolyhazy1966a}, for example, may result in modifications of the Schr\"odinger equation, decoherence and a quantum-classical transition~\cite{Bassi2017a}. Identifying such modifications would provide new insights for understanding the fundamental laws of Nature. With the space-based platform MAQRO proposed here, these questions will be answered by observing free quantum evolution and interference of dielectric test particles with radii of about $100\,$nm. 

Quantum physics predicts that physical systems of arbitrary size and complexity can be in superpositions of distinct states. This is well illustrated by Schr\"{o}dinger's dead-and-alive cat gedankenexperiment~\cite{Schroedinger1935a}. While challenging our intuitive understanding of reality, the predictions of quantum theory have been confirmed with large molecules consisting of  thousands of atoms and masses of up to $3\times 10^4\,$amu (atomic mass units)~\cite{Fein2019a}.

MAQRO will investigate sources of decoherence affecting macroscopic quantum superpositions, such as the scattering of residual gas, or solar/cosmic radiation. MAQRO also has the potential to detect some forms of dark or exotic matter~\cite{Riedel2013a,Bateman2015a,Riedel2017a}, and for rare scattering processes in a space environment. MAQRO could provide experimental input for the standard model of cosmology, for possible extensions of the standard model of particle physics, and for a better understanding of the origin of the Universe and the foundations of physics.
%

\subsection{\label{subsec::ScienceObjectives}Science Objectives} 
The science objectives (SOs) of the medium-size mission proposal MAQRO submitted to the European Space Agency (ESA)~\cite{Kaltenbaek2012b,Kaltenbaek2016a} were as follows:
\setlist{nolistsep}
\begin{itemize}[noitemsep]
    \item \textbf{SO1}: Testing the predictions of quantum physics in parameter regimes that overlap with ground-based tests.
    \item \textbf{SO2}: Testing standard decoherence mechanisms with test particle sizes and masses beyond existing experiments.
    \item \textbf{SO3}: Testing gravitational decoherence with sufficiently massive test particles.
\end{itemize}
This research campaign will investigate whether the experiments suggested for MAQRO can be adapted to address additional objectives (AOs) including but not limited to:
\begin{itemize}[noitemsep]
    \item \textbf{AO1}: Measuring the effect of decoherence on rotational quantum revivals.
    \item \textbf{AO2}: Using the macroscopic quantum systems on MAQRO or trapped charged particles as highly-sensitive detectors for dark or exotic matter.
\end{itemize}

MAQRO is based on optomechanics with optically trapped dielectric particles~\cite{Chang2010a,RomeroIsart2010a,Barker2010a}. After their release, their free evolution is monitored over long periods of time. While this remains the central approach of MAQRO, we will also investigate the feasibility of measuring rotational quantum revivals as an additional experimental method (AO1 above).

Optically trapped particles may be able to detect dark matter~\cite{Monteiro2020a} and exotic physics, especially when operated at~\cite{Afek2021a,Carney:2020xol} or beyond~\cite{Riedel2013a,Bateman2015a,Riedel2017a} the standard quantum limit. Attaining the requirements for achieving the primary science objectives of MAQRO (SO1--3) will enhance the detection sensitivity to sources of anomalous diffusion~\cite{Riedel2015a,Branford2019a} to unprecedented degrees. 
The environmental isolation and large test particle masses in MAQRO provide perfect conditions for detecting impulses or accelerations imparted by relic dark matter particles over a wide range of parameter space for mass and interaction strength~\cite{Bateman2015a,Riedel2017a} (AO2 above). We also expect improved sensitivity to additional models of dark matter including composite particles~\cite{Monteiro2020a} or ultralight dark matter~\cite{Carney:2019cio}, which will be investigated in detail. 

\subsection{\label{subsec::ScienceRequirements}Science Requirements}
A macroscopic quantum object in superposition must be isolated from its environment to prevent decoherence by scattering of surrounding particles. Depending on how much information a scattering event carries away, the decoherence will be (a) in the short-wavelength limit, where a single scattering event may destroy the superposition, or (b) in the long-wavelength limit, where the superposition may survive many scattering events~\cite{Joos1985a,Schlosshauer2007a}. To observe macroscopic superpositions, the probability of short-wavelength scattering events must be negligible~\cite{Voirin2019a}. This can be monitored using weakly trapped test particles or their free evolution. In the presence of long-wavelength decoherence, the evolution of a test particle's center-of-mass (CM) in 1D is given by a Markovian master equation\cite{Schlosshauer2007a}:
\setlength{\abovedisplayskip}{3pt}
\setlength{\belowdisplayskip}{3pt}
\begin{equation}
\label{eq::evolution}\dot{\hat{\rho}}(t) = (i/2 m \hbar) [\hat{\rho},\hat{p}^2] - \Lambda [\hat{\mathcal{O}},[\hat{\mathcal{O}},\hat{\rho}]].
\end{equation}
$\hbar$: Planck's constant, $m$: test-particle mass, dot: time derivative, $\hat{x},\hat{p}$: position and momentum operators; $\hat{\rho}(t)$: density operator at time $t$. $\hat{\mathcal{O}}$ can be replaced by $\hat{x}$ for decoherence in the position basis, or by $\hat{p}^2/(2 m)$ for decoherence in the energy basis, such as predicted by theories in the frameworks of general relativity and quantum field theory~\cite{Blencowe2013a,Anastopoulos2021a}. The two terms on the right-hand side represent coherent quantum evolution and decoherence, respectively. The decoherence parameter $\Lambda$ encodes the strength of both environmental decoherence and of fundamental deviations from the predictions of quantum physics. 

We will describe the science requirements needed to fulfil the science objectives in terms of the range of values of $\Lambda$ our experiments need to be sensitive to. Based on this, we will derive the required test masses, particle sizes, particle and environment temperatures, and vacuum conditions. To illustrate this for testing gravitational decoherence~\cite{Bassi2017a} with transparent, dielectric particles, we will consider the ``K model'' of gravitational decoherence by K\'{a}rolyh\'{a}zy~\cite{Karolyhazy1966a,Frenkel1990a} and the ``DP model'' of gravitationally induced collapse by Di\'{o}si and Penrose~\cite{Diosi1987a,Penrose1996a,Diosi2007a,Penrose2014a} for continuous mass distributions~\cite{Kaltenbaek2021a}. For particles below a critical size, the K model predicts a negligibly small $\Lambda$. Close to that critical size ($\sim 100\,$nm), the predictions of the two models intersect. In MAQRO, we aim to achieve sensitivity to the values of $\Lambda$ in this range or larger. This corresponds to radii of $100-180\,$nm, and a mass of $\sim 10^{10}\,$amu. Sensitivity in this regime requires the detection of decoherence parameters of $\Lambda \ge 10^{11}\,\mathrm{s^{-1}m^{-2}}$. For a superposition size $\Delta x$ comparable to the particle radius, this requires coherence times $1/(\Lambda \Delta x^2) \ge 10^2\,$s. Reducing these times would require even larger superpositions, which become increasingly difficult to achieve. Free evolution times of a few $100\,$s may be feasible, depending on the attainable vacuum and on the shielding from electrons~\cite{Voirin2019a}. For example, a measurement time of $\sim 100\,$s will require a scattering rate below $10~$mHz. In helium gas at $20\,$K, that value corresponds to $\lesssim 10^{-15}$mbar. Achieving these extremely high vacuum (XHV) conditions is a critical challenge~\cite{Voirin2019a}. An alternative approach may be to apply the method of quantum mechanical squeezing. This could reduce the required free-fall times~\cite{Branford2019a} with the downsides of adding complexity and noise. Equivalently, if we had an XHV environment, squeezing would enable tests of even weaker decoherence effects.

MAQRO will measure $\Lambda$ using dielectric test particles of varying transparent materials and radii via the following methods~\cite{Kaltenbaek2016a,Voirin2019a}:
\setlist{nolistsep}
\begin{enumerate}[label={(\alph*)},noitemsep]
    \item monitoring the heating of the CM motion of a weakly trapped particle.
    \item monitoring the wavepacket expansion of particles released from a trap.
    \item \label{enum::exp:Interference} observing near-field matter-wave interference.
\end{enumerate}
We will investigate the feasibility of integrating further measurement techniques:
\setlist{nolistsep}
\begin{enumerate}[label={(\alph*)},start=4,noitemsep]
\item orientational quantum revivals of rotating test particles.
\item monitoring trapped charged particles to detect dark or exotic matter.
\end{enumerate}
To define the science requirements, we will focus on method \ref{enum::exp:Interference}. For near-field interferometry with a grating period $d$, the Talbot time $m d^2/(2\pi \hbar)$ determines the time scale. For $10^{10}\,$amu test particles and a grating period of $100\,$nm, this yields a baseline value of $100\,$s~\cite{Kaltenbaek2016a,Gasbarri2021a}.

To achieve a sensitivity to values of $\Lambda$ as low as $10^{11}\,\mathrm{s^{-1}m^{-2}}$, decoherence effects have to be suppressed very efficiently, which means that the requirements on both environment and test particle become very stringent, for example in terms of temperature. The precise limits on these temperatures depend on the material properties of the test particles. For silica particles, the limit is $\lesssim 20\,$K for both particle and environment. For silicon particles, the requirements remain more relaxed with temperatures $\lesssim 50\,$K.

\subsection{\label{subsec::CaseForSpace}The Case for Space}
Experiments testing macroscopic quantum superpositions in space have several key advantages. Some objectives may not be achievable on ground:
\setlist{nolistsep}
\begin{enumerate}[label={(\alph*)},start=1,noitemsep]
\item \label{enum::CStimes}\textit{Long coherence times and free-evolution times}.\\
Observing the evolution of macroscopic quantum states on Earth requires trapping test particles, e.g., via optical~\cite{Chang2010a,RomeroIsart2010a,Barker2010a}, electrostatic~\cite{Martinetz2020a} or magnetic fields~\cite{Pino2018a}. Trapping will inevitably couple the systems to vibrations or to noise in the trapping potential or lead to decoherence due to scattering or absorption~\cite{RomeroIsart2011b,Nimmrichter2014a}. Methods to accelerate the time evolution of the quantum state~\cite{Pino2018a} may add excess noise.
\item \label{enum::CSvibrations}\textit{Isolation from low-frequency vibration or Newtonian noise}.\\
Such noise can, e.g., wash out interference patterns or heat the CM motion of trapped particles. Space can provide excellent microgravity ($\mu$g) conditions~\cite{armano2016a}.
\item \label{enum::CSgravity}\textit{Avoiding dephasing in gravitational potentials}.\\
Gravitational time dilation can lead to dephasing between different branches of superpositions in a gravitational field~\cite{Pikovski2015a,Zych2017a}. While this is not an issue for freely falling interfering particles, it may become relevant if guiding potentials are employed~\cite{Roura2020}.
\item \label{enum::CSexotic}\textit{Avoiding the shielding of dark matter/exotic matter by the atmosphere}.\\
Some dark and exotic matter candidates within the detection range of MAQRO would be blocked by, or thermalize with, Earth's atmosphere before reaching terrestrial detectors~\cite{Bateman2015a,Riedel2017a}. MAQRO, in space,  would have clean exposure to any dark matter flux coming from outside the solar system, including the anisotropic dark matter ``wind'' that would give a directional signal~\cite{Riedel2017a}.
\end{enumerate}

\section{\label{sec::MissionDesign}Mission Design and Technological Readiness}
MAQRO can harness a space environment for experiments at cryogenic temperatures and in XHV on an external optical bench~\cite{Kaltenbaek2012b,Kaltenbaek2016a}. While earlier studies aimed to achieve the required temperature via passive radiative cooling~\cite{PilanZanoni2016a}, the ``quantum physics platform'' (QPPF) feasibility study~\cite{Voirin2019a} proposed additional active cooling to ensure temperatures $< 20\,$K, and to encase the optical bench, which will prevent electrons from charging the test particles. The cover, however, renders XHV more difficult.

For a successful realization of MAQRO, three \textit{critical issues} (C1-C3) will have to be addressed~\cite{Voirin2019a}. An XHV environment will need to be established and the test particles will need to be protected from electrons (C1). These particles need to be loaded into an optical trap, and their charge, mass, radius and material properties must be well characterized (C2). To implement matter-wave interferometry, a phase grating will be used to prepare macroscopic superpositions~\cite{Bateman2014a,Kaltenbaek2016a}. For large particles, the scattering of grating photons may decohere the quantum state (C3).

This needs to be done while fulfilling other key requirements including cryogenic temperatures. Developing relevant solutions will be an essential part of this campaign. Experiments on the ground and pathfinders in space will address C1. With respect to C2, groups continue working on reliable methods to load test particles into optical traps in vacuum (e.g., see~\cite{Bykov2019a,Nikkhou2021a}). Ideas based on QPPF~\cite{Voirin2019a} and on an ESA-funded study~\cite{Schmid2014a} are being investigated by ESA contractors. With respect to C3, it was shown that matter-wave interferometry is still possible even for large test particles~\cite{Gasbarri2021a}. More work is required to ensure MAQRO's science objectives will be achieved, or to develop an alternative approach for preparing macroscopic CM superpositions.

\begin{figure}[h!]
  \includegraphics[width=0.65\linewidth]{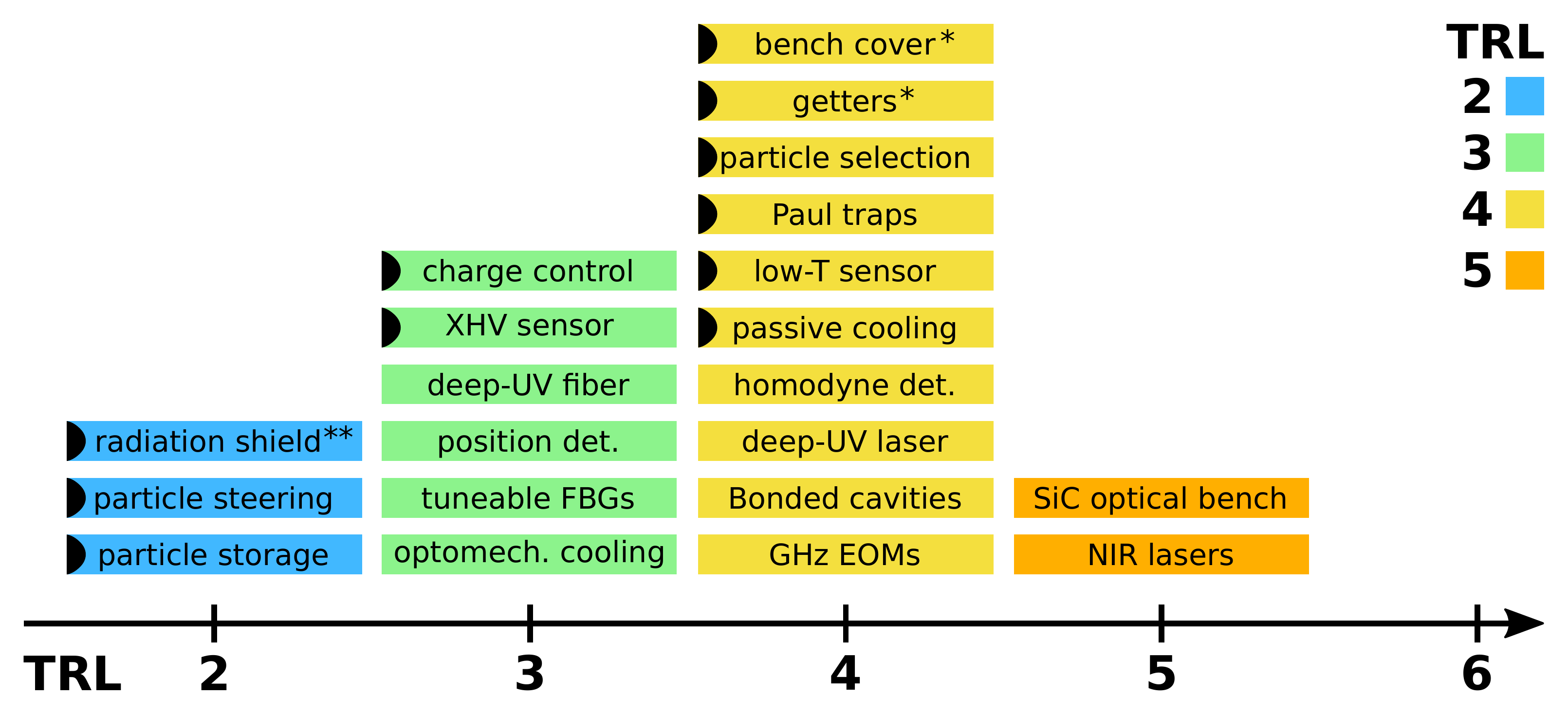}
  \caption{\label{fig::TRLs}\csentence{Overview of technologies with a TRL $< 6$.} Black semicircles: critical. *: required for a covered optical bench (OB). **: required for an OB open to space.}
\end{figure}

The TRL of key technologies needs to be increased, as shown in Fig.~\ref{fig::TRLs}: (1) optomechanical cooling, (2) electro-optic modulators (EOMs), (3) near-infrared (NIR) lasers, (4) open-access cavities, (5) position detection, (6) a deep-UV laser source, (7) fibers for deep UV, (8) fiber Bragg gratings (FBGs), (9) homodyne detection, (10) a silicon-carbide (SiC) optical bench, (11) passive radiative cooling, (12) XHV pressure sensors, (13) low-temperature (low-T) sensors, (14) particle charge control, (15) Paul traps, (16) particle characterization, (17) storage, and (18) steering. If the optical bench is covered, the cover (19) and non-evaporable getters (20) have to be developed. If the optical bench is open to space, electromagnetic shielding (21) must be developed to protect the test particles from electrons.

\section{\label{sec::ResearchCampaign}The Research Campaign}
Technology development during the current decade could enable the implementation of MAQRO before 2040. We envisage in-orbit demonstrations to de-risk the critical components identified above and leveraging small satellite systems to accelerate mission development~\cite{oi2017nanosatellites}.

\begin{figure}[h!]
  \includegraphics[width=0.65\linewidth]{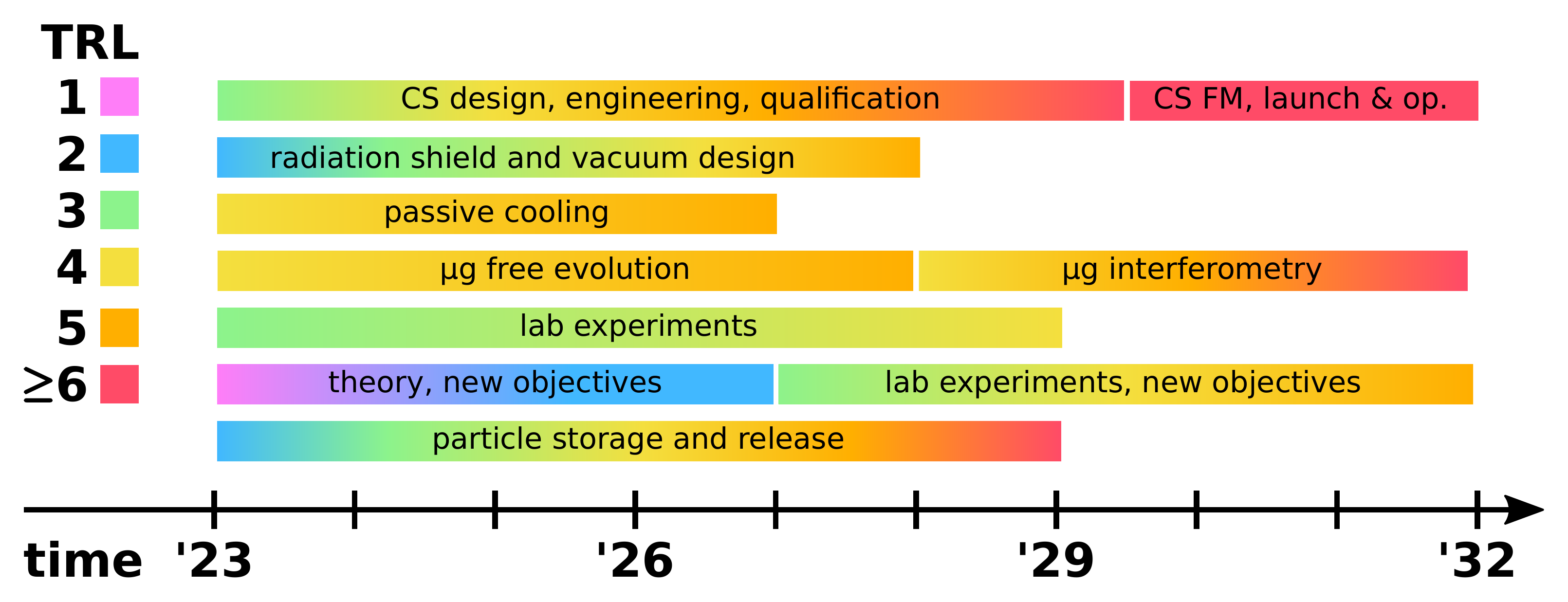}
  \caption{\label{fig::timeline}\csentence{Timeline of key activities.} CS: CubeSat, FM: flight model, op.: operation.}
\end{figure}

Fig.~\ref{fig::timeline} provides an overview of key activities of this research campaign:
\begin{itemize}[noitemsep]
\item theoretical analysis of new science objectives, the corresponding scientific requirements, and the feasibility of addressing these objectives with MAQRO.
\item laboratory experiments on matter-wave interferometry, rotational revivals, and proof-of-principle tests addressing potential new science objectives.
\item design and test passive cooling for MAQRO and CubeSat pathfinders.
\item design and test radiation shields for MAQRO and CubeSat pathfinders to protect the test particles from electrons on platforms open to space.
\item experiments on free quantum evolution in $\mu$g may reach TRL 3 by 2023. These will be followed by matter-wave interferometry in $\mu$g.
\item several methods for particle storage, release and characterization are under investigation and may reach TRL 3 by the time this campaign starts.
\end{itemize}

Experiments will aim to close the gap between state-of-the-art ($\sim 3\times 10^4\,$amu~\cite{Fein2019a}) and the minimum test masses in MAQRO ($\sim 10^8\,$amu) with CM interference~\cite{Bateman2014a,Kialka2021a}, orientational quantum revivals~\cite{Stickler2018a,Schaefer2021a} or electrostatically levitated objects~\cite{Martinetz2020a,Penny2021a}.

\subsection{MAQRO Pathfinders}
$\mu$g and space experiments will be essential to better understand the effects of a space environment on macroscopic quantum superpositions, and how to achieve XHV and protection from radiation. This will be addressed by CubeSat pathfinders:
\begin{itemize}
    \item MAQRO-PF1 LEO CubeSat: demonstrator for key technologies, e.g. passive-cooling concepts~\cite{PilanZanoni2016a,SierraLobo2015a} or trapped particles as XHV sensors after a wakeshield~\cite{Wuenscher1970a,melfi1976molecular,strozier2001wake}. 
    \item MAQRO-PF2 CubeSat Ride-share to L2: Test XHV, radiation conditions \& passive cooling. Show long free evolution for macroscopic superpositions.
\end{itemize}

\subsection{Cost Estimate}
We estimate the cost at completion of MAQRO as $\sim\$550$M~\cite{Kaltenbaek2012b,Kaltenbaek2016a,Voirin2019a}, not taking into account remaining research and development. The wet mass of MAQRO is 1700-1900~kg~\cite{Kaltenbaek2016a,Voirin2019a}. An orbit around the Earth-Sun Lagrange point L2 is optimal~\cite{Voirin2019a}.

In addition, we assume $\$50$M for technology development, instrument delivery, spacecraft I\&T, and missions operations support. We estimate $\$120$M for the research campaign during this decade. MAQRO-PF2 can be executed for an estimated $\$15$M. The cost and schedule information in this document is of a budgetary and planning nature. It does not constitute a commitment on the part of any of the authors or their host institutions. We encourage a joint ESA-NASA effort to advance MAQRO, in the spirit of BECCAL and other multi-national scientific collaborations. Parallel support would multiply the science return, with this effort leveraging the expertise of partners across Europe and the US.

\section{\label{sec::Summary}Summary}
Experiments and theoretical studies performed in the current campaign will address remaining critical issues in MAQRO and investigate the feasibility of adding novel experimental methods and science objectives. Developing key technologies will lay the groundwork for pathfinder CubeSat missions in LEO and around the Earth-Sun Lagrange point L2. These will act as in-orbit demonstrators and provide insights into how the space environment affects quantum systems, and how it can be harnessed for realizing macroscopic quantum experiments.
\clearpage
\bibliographystyle{bmc-mathphys}
\bibliography{MAQROreferences}

\end{document}